\documentstyle[12pt,aasms4]{article}

\def\etal{{\it et al.}}

\def\eg{{\it e.g.}}
\def\lap{\hbox{${_{\displaystyle<}\atop^{\displaystyle\sim}}$}}
\def\gap{\hbox{${_{\displaystyle>}\atop^{\displaystyle\sim}}$}}

\begin{document}

\title{EVIDENCE FOR BRIGHTNESS-DEPENDENT ANISOTROPY OF GAMMA RAY
BURSTS AND ITS GALACTIC INTERPRETATION}
\author{Bennett Link}
\affil{Department of Physics, Montana State University, Bozeman,
MT 59717; Los Alamos National Laboratory; blink@dante.physics.montana.edu}
\and
\author{Richard I. Epstein} 
\affil{Los Alamos National Laboratory, Mail Stop D436, Los
Alamos, NM 87545; Department of Physics
and Astronomy, University of New Mexico; epstein@lanl.gov}

\begin{abstract}

We show that the brighter gamma-ray bursts (GRBs) in the BATSE 3B
catalog concentrate slightly toward the Galactic plane and center,
suggesting that at least some bursts originate within the Galaxy.  To
develop an interpretation of this {\it brightness-dependent
anisotropy}, we consider GRBs distributed in a thick disk centered on
the Galaxy.  As an approximation to the true luminosity
distribution, we divide bursts into two distinct luminosity classes.
Most bursts originate from low-luminosity, nearby sources while the
relatively few high-luminosity sources trace the shape of the thick
disk.  We find that characteristic disk dimensions as small as 20 kpc
in thickness and 30 kpc in radial extent can match the observed
brightness-dependent anisotropy as well as the number-flux
distribution; these dimensions are smaller than previously considered
viable. For bursts distributed in a disk of these compact dimensions,
the low-luminosity sources must have peak powers of $\lap 5\times
10^{40}$ erg s$^{-1}$ and rate densities of $\gap 5\times 10^{-3}$
yr$^{-1}$ kpc$^{-3}$. The high-luminosity sources must have a peak
luminosity of $\simeq 10^{42}$ erg s$^{-1}$ but their rate density
need be only 5\% or less than that of the low-luminosity sources.  If
the Andromeda Galaxy contained a similar population of high-luminosity
sources, their peak gamma-ray fluxes would be less than $\sim 0.1$
s$^{-1}$ cm$^{-2}$, making their detection problematic.  Statistical
analyses of the gamma-ray burst events obtained since the compilation
of the 3B catalog will determine whether the thick-disk interpretation
is preferred over the isotropic one.

\end{abstract}

\section{INTRODUCTION}

The 3B catalog (\cite{3Bcat}) of gamma ray bursts (GRBs) of the Burst
and Transient Source Experiment (BATSE) on board the {\em Compton
Gamma Ray Observatory} shows that the overall burst distribution is
highly isotropic, suggesting a cosmological origin.  The mean values
of the Galactic quadrupole and dipole moments of this burst
distribution are consistent, within better than one standard
deviation, with the events being isotropic on the sky.  These
statistics, however, are strongly weighted by the numerous faint GRBs
not much brighter than the detection threshold.  The overall isotropy
does not preclude the possibility that GRB subsets exhibit anisotropy,
which would be indicative of a Galactic origin.  Atteia
\& Dezalay (1993) present evidence for  brightness-dependent anisotropy,
and the study of  Briggs \etal\ (1996) shows hints of such a 
correlation. 

Brightness-dependent anisotropy would arise if GRBs originated from a
Galactic distribution of sources, occurring with a range of
luminosities. The bursters of lowest luminosity would be detected from
relatively near the sun and could generate the numerous, faint
isotropic events. The high-luminosity ones could be visible at the
extremities of the galaxy and appear as anisotropic bright events.
Equivalently, there could be one class of bursters whose emission is
strongly beamed so that the bursters are faint when viewed from most
directions but bright when viewed along the beam.  Mao \& Paczy\'nski
(1992), Higdon \& Lingenfelter (1992), Smith \& Lamb (1993) and Smith
(1995) have investigated two-luminosity class Galactic GRB models. 

In this Letter we describe the observational evidence for
brightness-dependent anisotropy and construct two-component Galactic
models which are consistent with the observations.  We show that
brighter bursts in the BATSE 3B catalog tend to concentrate toward the
Galactic plane, and to a lesser extent toward the Galactic center.  We
then develop a class of thick-disk models which match the angular
data, as well as the observed number of bursts per peak flux interval.

\section{DATA ANALYSIS}

The BATSE 3B catalog  contains data for 1122 GRBs. 
To search for  brightness-dependent anisotropy we use the subset
of 847 bursts for which both peak fluxes and positions have been
determined. For this sample we evaluate the cumulative Galactic
dipole  and quadrupole moments:
\begin{equation}
\langle \cos \theta \rangle_P = {1\over N_P} \sum_{P_i  \geq
P}\cos \theta_i, 
\end{equation} 
\begin{equation}
\langle  \sin ^2 b -{1 \over 3} \rangle_P = {1\over N_P}
\sum_{P_i  \geq P} (
\sin ^2 b_i -{1 \over 3}). 
\end{equation} 
In these expressions $P_i$ is the peak number flux of
the $i$th burst, 
$b_i$  is its Galactic latitude, and
$\theta_i$  is the angle between the event and the Galactic
center. The averages are taken over the $N_P$ bursts with peak
fluxes of $P$ or greater. 

The jagged solid lines in Fig. 1 show the cumulative dipole and
quadrupole moments as functions of peak flux. The peak fluxes were
measured at 64 ms resolution. For an isotropic distribution of events, 
both moments would be zero. However, because the BATSE sky coverage
was uneven, the expected values of the moments for an isotropic
distribution are $\langle \cos
\theta
\rangle _{\rm iso}= -0.013$ and $\langle  \sin ^2 b -{1 \over 3}
\rangle _{\rm iso} =  -0.005$ (\cite{3Bcat}). These expected values are
indicated by the  solid horizontal lines.  
The positive values  of the cumulative dipole
moment for the  brighter gamma-ray bursts suggest that these
events are somewhat clustered in the direction of the Galactic center, while
the negative values of the cumulative quadrupole moment for the
brighter events indicate a preference for the Galactic
plane.

To assess the significance of the observed deviations from the
isotropic values, we compute the {\it statistical residuals}, $\chi_D$
and $\chi_Q$, for the dipole and quadrupole moments, defined as 
\begin{equation}
\chi_D \equiv [\langle \cos \theta \rangle_P - 
\langle \cos \theta \rangle _{\rm iso}]/\sigma_D, 
\end{equation}
\begin{equation}
\chi_Q \equiv [\langle \sin^2 b \rangle_P - 
\langle \sin^2 b \rangle _{\rm iso}]/\sigma_Q, 
\end{equation}
where the $\sigma_i$ are the standard deviations.  For a
sample of $N_P$ isotropically distributed bursts, the standard
deviations are $\sigma_D = \sqrt{ 1/3 N_P} $ and $\sigma_Q =\sqrt{4/45
N_P}$. The solid lines in Fig. 2 show the statistical residuals
between the BATSE data and an isotropic source distribution.  The
magnitude of $\chi_Q$ exceeds unity for peak fluxes below 
$\sim 15$ s$^{-1}$ cm$^{-2}$, and it reaches $\sim 3$ at $P \sim
5$ s$^{-1}$ cm$^{-2}$.  This concentration of bright bursts toward the
Galactic plane suggests, but does not prove, that 
the isotropic interpretation is inadequate. 
The evidence for a concentration in the
direction of the Galactic center is not as strong.  The values of
$\chi_D$ exceed unity for peak fluxes between $P \sim 4\ {\rm and}\
15 $ s$^{-1}$ cm$^{-2}$, but do not go beyond $\sim 2$ for any flux. 
Figure 3 shows the differential distribution, $ dN_P/d \ln
P$. The deviation from a slope of -3/2 below a peak flux of $\sim 10$
s$^{-1}$ cm$^{-2}$ suggests inhomogeneity in the source
distribution. Below a peak flux of $\sim 1.5$ s$^{-1}$ cm$^{-2}$,
the effects of detector efficiency become significant.  

\section{CALCULATIONS}

To illustrate that a galactic interpretation of GRBs is
consistent with the observed angular and flux distributions, 
we consider GRBs distributed in a thick disk centered on
the Galaxy.  As an approximation to the true luminosity
distribution, we divide bursts into two distinct classes
with luminosities $L_1$ and $L_2$ ($L_1>L_2$). The GRB {\em rate-densities}, 
$\dot n_i$ (yr$^{-1}$
kpc$^{-3}$), are represented by an oblate ellipsoid with a Gaussian
profile: 
\begin{equation}
\dot n_i  =
\dot{\cal{N}}_i 
\exp \left[ -{1\over 2}\left({ \rho \over H_\rho} \right)^2 -
            {1\over 2}\left({ z\over H_z} \right)^2 \right ]; \qquad i=1,2, 
\end{equation}
where $\rho$ is the polar radial distance from the Galactic center, 
$z$ is the distance from the Galactic plane, and 
$\dot{\cal{N}}_1$ and $\dot{\cal{N}}_2$ represent the central rate 
densities for the high- and low-luminosity sources. 

We choose $L_2$ sufficiently small that the low-luminosity
component is visible only to distances much less than the 
disk thickness, and thus appears to have a uniform
density. An upper bound on $L_2$ is found by requiring that the
low-luminosity bursters near the edge of the source ellipsoid
produce events that are well below a peak flux of 
$\sim 1$ s$^{-1}$ cm$^{-2}$. We take the rate at which photons
are emitted in the BATSE trigger window (50-300 keV) for each source 
as $L_i/\epsilon$, where $\epsilon\simeq 300$ keV. 
The implied limit on
the power of the low-luminosity component is then
\begin{equation}
L_2 \ll 5 \times 10^{40} \left ({H_z\over 20\ {\rm kpc}}\right )^2\ {\rm erg\ s}^{-1},
\label{L_2_limit}
\end{equation}
where we have assumed spherically-symmetric
emission. The fiducial normalization
of equation
(4) was chosen because,   as we show below,  acceptable models have values of
$H_z$ as small as 20 kpc.

The rate at which bursts occur with a peak energy flux exceeding $P$
is 
\begin{equation}
\dot N (P) = \int_0^\pi d\theta\,\sin\theta \int_0^{2\pi} d\phi
                      \int_0^{d_1 (P)} dr\,r^2\,
             \dot n_1 (\rho, z) +
             {\dot{\cal{N}}_2\over 6 \pi^{1/2}}
             \left ({L_2\over \epsilon P}\right )^{3/2},
\label{rate}
\end{equation}
where $\phi$ is the angle of the source around the  Galactic center
measured from the Galactic plane, $r$ is
the distance from the observer to the source, and $d_1 (P)\equiv
(L_1/4\pi \epsilon P)^{1/2}$ is the
distance to a high-luminosity source  with observed flux
$P$. The
distances $\rho$ and $z$ are related to $\theta$ and $\phi$ by 
\begin{equation} 
\rho^2 + z^2 = R_0^2 + r^2 - 2R_0r\cos\theta; \qquad z = r\sin\phi\sin\theta,
\end{equation}
where $R_0=8.5$ kpc is the assumed distance to the Galactic center.

The second term in Eq. [\ref{rate}] represents the homogeneous, isotropic
contribution from the low-luminosity bursters. It gives
-3/2  logarithmic slope, while the high-luminosity bursters produce  deviations
from this behavior at low flux. 
In terms of these variables, the dipole  moment for a
sample
of bursts with fluxes exceeding
$P$ is
\begin{equation}
\langle\cos\theta\rangle_P  =
{1\over\dot N (P)} \int_0^\pi d\theta\,\sin\theta\cos\theta
 \int_0^{2\pi} d\phi \int_0^{d_1 (P)} dr\,r^2\,\dot n_1 (\rho, z),
\label{dipolemoment}
\end{equation}
and 
\begin{equation}
\langle\sin^2 b\rangle_P  =
{1\over\dot N (P)}\left ( \int_0^\pi d\theta\,\sin^3\theta\sin^2\phi
 \int_0^{2\pi} d\phi \int_0^{d_1 (P)} dr\,r^2\,\dot n_1 (\rho, z) +
       {\dot{\cal N}_2\over 18 \pi^{1/2}} 
        \left [ {L_2\over\epsilon P}\right ]^{3/2}
           \right ). 
\label{quadmoment}
\end{equation}
For $\dot{\cal{N}}_2\gg\dot{\cal{N}}_1$, the quadrupole moment $[\langle\sin^2
b\rangle_P - 1/3]$ vanishes. 

This model has five adjustable parameters: $H_\rho$, $H_z$, $L_1$, 
$\dot{\cal N}_1$ and $\dot{\cal N}_2 L_2^{3/2}$. We have searched 
this five dimensional
parameter space for models that are consistent with the 
observed angular and number-flux
distributions. Our consistency criteria are:

\begin{itemize}

\item The $\chi^2$ difference between the model differential
distribution ($dN_P/d\ln P$) and the observed distribution must be
less than 1.5 per degree of freedom.\footnote{For the two bins of
lowest peak flux, the instrument efficiency corrections significantly
affect the estimated number. Because these corrections are not
well-determined, we take the half range between the highest corrected
number and the lowest uncorrected number as the standard deviation,
with the center of this range as the most likely value.} 

\item The model's dipole and quadrupole moments must not differ from
the data by more than 1.5 standard deviations for any observed peak
flux.

\end{itemize}

Figure 5 shows the region in the $H_\rho$-$H_z$ plane where acceptable
models are found. The acceptable models are thick disks with aspect
ratios $H_\rho/H_z \sim 1.5$. The smallest allowed models, which we
call {\em compact models}, have dimensions of $H_\rho \sim 30$ kpc
and $H_z \sim 20$ kpc.  Figs. 1-4 show the agreement of a compact
model with the angular and flux distributions. In these compact models
the high-luminosity component has a peak luminosity of $L_1 \simeq
10^{42}$ erg s$^{-1}$ and a central rate-density factor of $\dot{\cal N}_1
\sim 3 \times 10^{-4}$ yr$^{-1}$ kpc$^{-3}$.\footnote{The 
normalization of the full sky burst rate was fixed to match the
cumulative rate of 550 yr$^{-1}$ at $P=1$ cm$^{-2}$ s$^{-1}$
determined from the 1B catalog (\cite{1B}).} The low-luminosity
component is characterized by $(\dot{\cal N}_2/10^{-2}\ {\rm yr}^{-1}
{\rm kpc}^{-3}) (L_2/10^{40}\ {\rm erg \ s}^{-1} )^{3/2} \simeq 6$.
Combining this relationship with the limit of Eq. (\ref{L_2_limit})
gives $\dot{\cal N}_2 \gg 5 \times 10^{-3}$ yr$^{-1}$ kpc$^{-3} \,
(H_z/ 20 {\rm kpc})^{-3}$.  This rate-density can be accommodated by a
population of Galactic neutron stars that fill the thick disk and
occasionally produce gamma-ray bursts.  For example, if the thick disk
contained $10^8$ neutron stars each of which produced a low-luminosity
burst approximately every $t_{\rm rep}$ years, the central
rate-density would be $\dot{\cal N}_2\sim 350 / t_{\rm rep}$
yr$^{-1}$ kpc$^{-3}\, (H_z/ 20 {\rm kpc})^{-3}$, if we assume the
aspect ratio is $H_\rho/H_z = 1.5$.  Consistency with the observed
gamma-ray burst rate requires $t_{\rm rep} \ll 7 \times 10^4$
yr, an acceptable bound for burst repetitions.

The dashed lines in Fig. 1 show the cumulative moments of the compact
models; the offsets due to incomplete sky coverage ($\langle \cos\theta
\rangle _{\rm iso}$, $\langle  \sin ^2 b -{1 \over 3}
\rangle _{\rm iso}$) have been added. 
The dotted lines in Fig. 2 give the statistical residuals between the
data and the compact model results. The selection process for the
acceptable models ensures that these residuals are smaller than those
for the isotropic hypothesis. Figure 3 shows that the compact model
yields a number-peak flux distribution that is consistent with the
BATSE data.  At the faint end, the model falls far below the corrected
BATSE data point.  However, because the BATSE detector efficiency is
not accurately known for the lowest peak fluxes (Meegan, private
communication), this discrepancy is not necessarily significant.
Figure 4 shows that the observed and predicted cumulative
distributions are also in good agreement, except for the
poorly-determined faint end of the distribution.

\section{SUMMARY AND CONCLUSIONS}

The BATSE 3B catalog shows indications of a brightness-dependent
anisotropy suggesting that the brighter sources cluster toward the
Galactic plane, and to a lesser extent, the Galactic center.  This
clustering, if a reflection of the true GRB angular distribution, is
inconsistent with the cosmological interpretation.  We show that the
observed angular and flux distributions are consistent with a Galactic
distribution of bursters residing in a thick disk, with a bimodal
luminosity function. Most bursts originate from low-luminosity, nearby
sources while the relatively few ($\lap 5$\%) high-luminosity sources
trace the shape of the thick disk.  The high-luminosity sources have
peak luminosities $\simeq 10^{42}$ ergs s$^{-1}$, and are at least 25
times more luminous than the low-luminosity sources. We find that the
dimensions of the disk can be as small as 20 kpc in thickness and 30
kpc in radial extent; this GRB source distribution is more compact 
than normally considered viable (see, \eg, \cite{isotropy}). If the Andromeda
Galaxy contained a similar population of high-luminosity sources,
their peak gamma-ray fluxes would be less than $\sim 0.1$ s$^{-1}$
cm$^{-2}$, making their detection problematic.

It would not be fruitful to attempt to use the data in the BATSE 3B
catalog to determine whether this two-component thick disk model is
preferred over the isotropic interpretation. Since our thick-disk
model was motivated by the 3B data, we do not attribute 
statistical significance to the agreement between the model and the
data.  On the other hand, data taken since the compilation of the 3B
catalog may be able to distinguish between these two interpretations.
As Fig. 1 shows, the expectations of the thick-disk and the isotropic
interpretations are quite different.  Only one or two years may be
sufficient to establish whether the trends of the two-component
thick-disk models are borne out by the new data.

\acknowledgements

It is a pleasure to thank C. Meegan for valuable discussions. 
This work was carried out under the auspices of the U. S.
Department of Energy and supported in part by IGPP at LANL, NASA
EPSCoR Grant \#291471 and a Compton Gamma Ray Observatory Guest
Investigator Grant.

\newpage

\figcaption{The cumulative Galactic moments as functions of peak
flux. The left panel shows the dipole moments and the right, the
quadrupole moments.  The solid lines show the observed
values, and the horizontal lines are the expected values for an
isotropic distribution of bursts.  The dotted lines  are expected
values for a  compact thick-disk model with $H_z =20$ kpc and
$H_\rho =30$ kpc.}

\figcaption{The statistical residuals of cumulative Galactic
moments as functions of peak flux.  The panels and lines are as
in Fig. 1.}

\figcaption{The differential number peak flux distribution, adapted
from Meegan \etal\ (1996, Fig. 10a). Shown is the number of bursts per
factor of 1.55 in peak flux, equivalent to 0.438$\,dN_p/d\ln P$.  The
upper error bars for the two lowest flux bins include corrections for
detector efficiency.  The curve represents a compact thick-disk model
with $H_z =20$ kpc and $H_\rho =30$ kpc. The dash-dotted line shows
the -3/2 power law expected for a homogeneous burst distribution.}

\figcaption{The cumulative number distribution. The thin solid line 
is the observed distribution, and 
the thick solid curve
is for a compact thick-disk model with $H_z =20$ kpc and $H_\rho =30$
kpc. The dotted line represents the contribution from low-luminosity
sources, while the dashed line is for the high-luminosity sources. }

\figcaption{The allowed region of the $H_\rho$-$H_z$ plane for
two-component Galactic gamma-ray bursts. }


\newpage


\begin{thebibliography}{}

\def\nature{{\rm Nature}}

\bibitem [Atteia \& Dezalay 1993]{AD93}Atteia, J. -L. \&
Dezalay, J. -P., 1993,
\aap, 274, L1.

\bibitem[Briggs \etal\ 1996]{isotropy} Briggs, M. S., Paciesas,
W. S.,  Pendleton, G. N., Meegan, C. A., Fishman, G. J., Horack,
J. M., Brock, M., Kouveliotou, C., Hartmann, D. H., \& Hakkila,
J. 1996, \apj, 459, 40.

\bibitem[Connors, Serlemitsos \& Swank  1996]{HEAO}Connors, A.,
Serlemitsos, P. J., \& Swank, J. H.  1986, \apj, 303, 769.

\bibitem[Fishman, \etal\ 1994]{1B}Fishman, G., \etal\ 1994, \apjs, 92,
229. 

\bibitem[Gotthelf, Hamilton \& Helfand 1996]{Einstein} Gotthelf,
E. V., Hamilton, T. T. \& Helfand, D. J., 1996, \apj, 466, 779.

\bibitem[Higdon  \& Lingenfelter 1992]{HL92}Higdon, J. C.  \&
Lingenfelter, R. E.,   1992,
\nature, 356, 132.

\bibitem[Higdon  \& Lingenfelter 1994]{HL94}Higdon, J. C.  \&
Lingenfelter, R. E.,   1994,
\apj, 434, 552.

\bibitem[Mao \& Paczy\'nski  1992]{MP92}Mao, S. \& Paczy\'nski,
B.  1992,
\apj, 389, L13.

\bibitem[Meegan 1997]{meegan} Meegan, C. A. 1997, {\sl private
communication}. 

\bibitem[Meegan \etal\ 1996]{3Bcat} Meegan, C. A. \etal , 1996,
\apjs, 106, 65.

\bibitem[Smith \& Lamb  1993]{SL93} Smith, I. A. \& Lamb, D.
Q.,  1993, \apj, 410, L23.

\bibitem[Smith   1995]{S95} Smith, I. A., 1993, \apj, 444, 686.


\end{thebibliography}
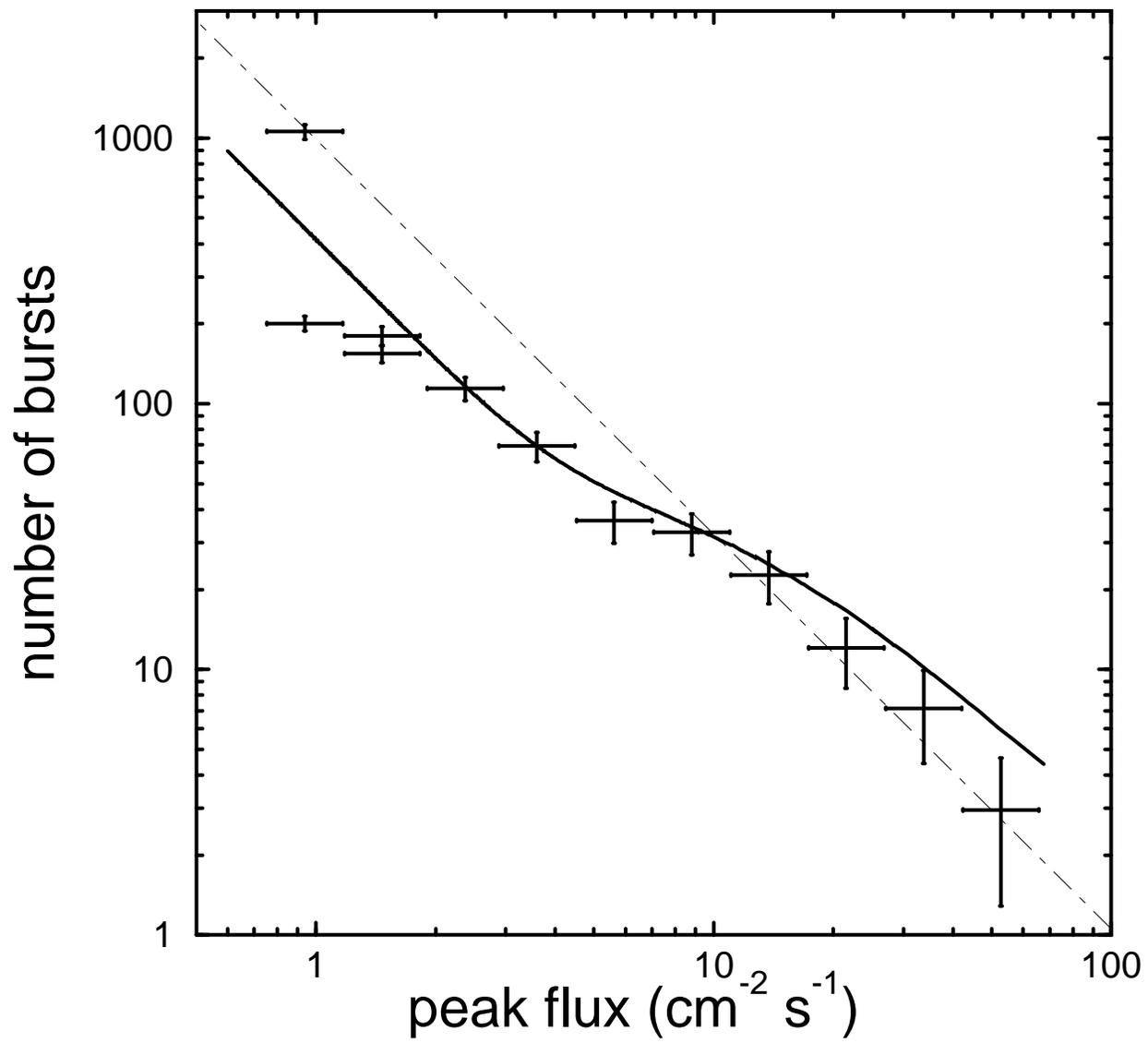
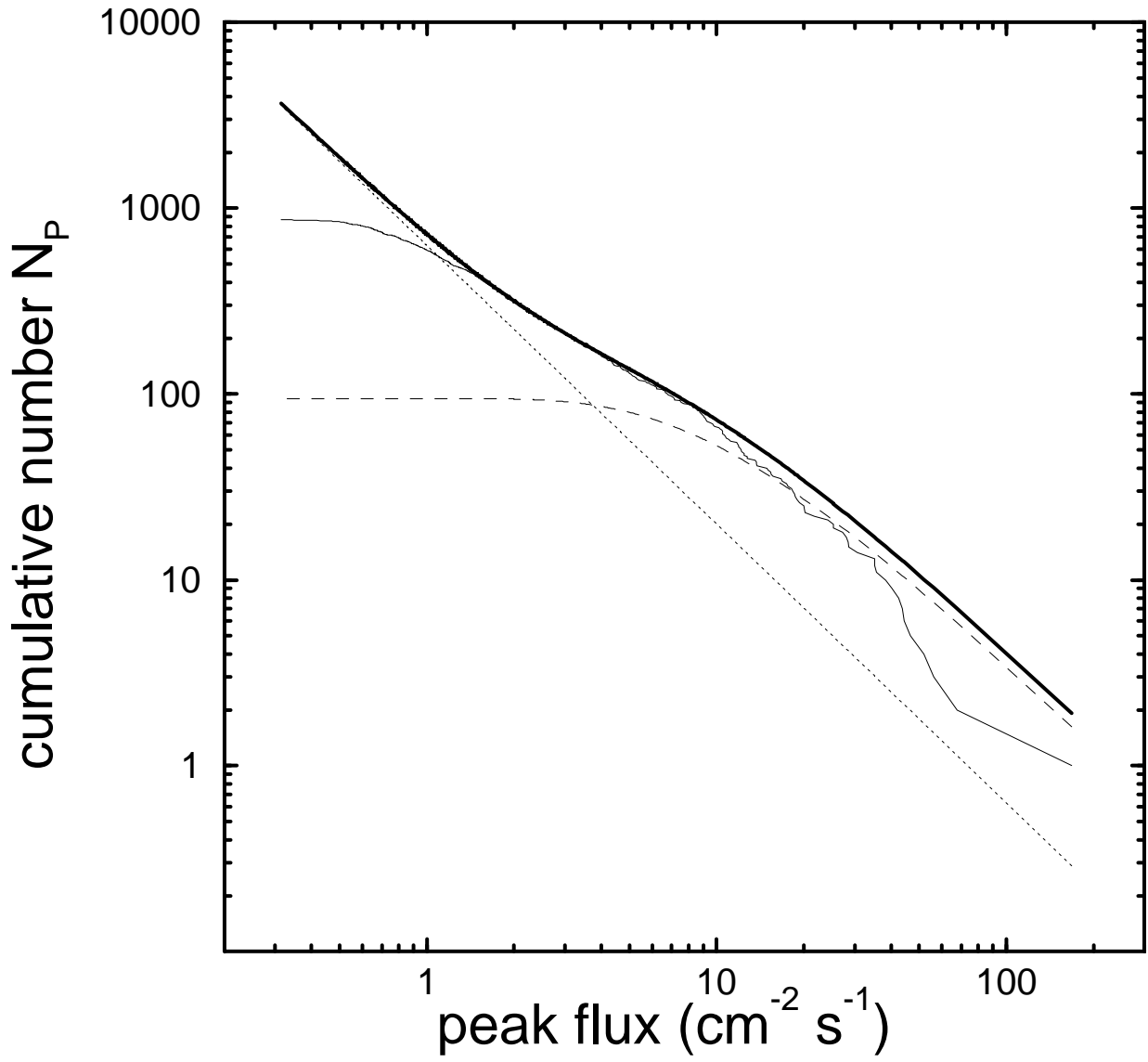
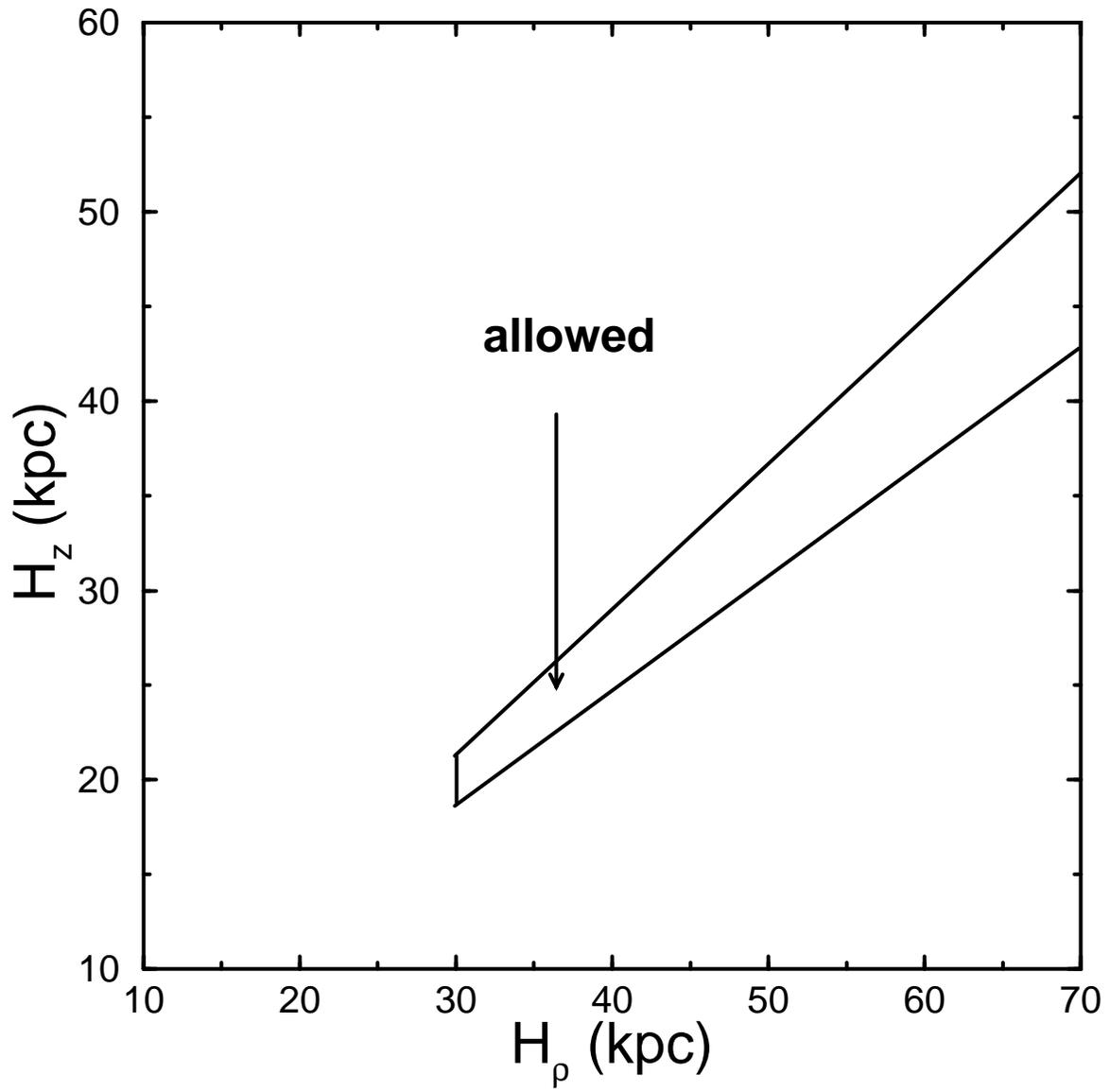
\end{document}